\documentclass[12pt,english]{article}
\usepackage[T1]{fontenc}
\usepackage[latin9]{inputenc}
\usepackage{geometry}
\usepackage[active]{srcltx}
\usepackage{amsmath}
\usepackage{amsthm}
\usepackage{amssymb}
\usepackage{stmaryrd}
\usepackage{graphicx}
\usepackage{esint}
\usepackage[numbers]{natbib}

\makeatletter
\newcommand{\lyxaddress}[1]{
	\par {\raggedright #1
	\vspace{1.4em}
	\noindent\par}
}

\def\vec#1{\boldsymbol{#1}}

\date{}

\makeatother

\usepackage{babel}
\begin{document}
\title{Self-contained two-layer shallow-water theory\\
of strong internal bores}
\author{J\={a}nis Priede$^{1,2}$}
\maketitle

\lyxaddress{$^{1}$Fluid and Complex Systems Research Centre, Coventry University,
\\
Coventry, CV1 5FB, United Kingdom\\
$^{2}$ Department of Physics, University of Latvia, Riga, LV-1004,
Latvia}
\begin{abstract}
We show that interfacial gravity waves comprising strong hydraulic
jumps (bores) can be described by a two-layer hydrostatic shallow-water
(SW) approximation without invoking additional front conditions. The
theory is based on a new SW momentum equation which is derived in
locally conservative form containing a free parameter $\alpha.$ This
parameter, which defines the relative contribution of each layer to
the pressure at the interface, affects only hydraulic jumps but not
continuous waves. The Rankine-Hugoniot jump conditions for the momentum
and mass conservation equations are found to be mathematically equivalent
to the classical front conditions, which were previously thought to
be outside the scope of SW approximation. Dimensional arguments suggest
that $\alpha$ depends on the density ratio. For nearly equal densities,
both layers are expected to affect interfacial pressure with approximately
equal weight coefficients, which corresponds to $\alpha\approx0.$
The front propagation velocity for $\alpha=0$ agrees well with experimental
and numerical results in a wide range of bore strengths. A remarkably
better agreement with high-accuracy numerical results is achieved
by $\alpha=\sqrt{5}-2,$ which yields the largest height that a stable
gravity current can have. 
\end{abstract}

\section{Introduction}

Shallow-water (SW) approximation is commonly used in the geophysical
fluid dynamics to model ocean currents and large-scale atmosphere
circulation \citep{Pedlosky1979}. Because such flows are typically
dominated by inertia and have a horizontal length scale much larger
than the characteristic depth, they can be treated as effectively
horizontal and vertically invariant. This simplifies the hydrodynamic
problem from three to two spatial dimensions, thus essentially reducing
computational complexity of such flows. The SW approximation can also
be used for modeling long gravity waves on the liquid surfaces or
interfaces in stably stratified fluid layers. The latter type of systems
are not only routinely used as simplified models of internal waves
in oceans \citep{Helfrich2006} but are also encountered in technological
applications like aluminum reduction cells \citep{Evans2007} and
the recently developed liquid metal batteries \citep{Kelley2018}.

In the commonly used hydrostatic SW approximation, waves are known
to become steeper with time and to develop vertical fronts analogous
to the shock waves in the gas dynamics \citep{Courant1948}. In the
fluid dynamics, such shocks are called hydraulic jumps or bores \citep{Stoker1958}
-- both terms are used interchangeably here. Hydraulic jumps can
also be present initially, for example, when fluid starts to flow
by breaking a dam or when a lock separating two liquids with different
densities is opened \citep{Esler2011}. Mathematically, hydraulic
jumps appear as discontinuities in the wave amplitude. Physically,
they encapsulate smooth variations of the flow field over the length
scales comparable to the layer depth.

It is commonly assumed that although the partial differential equations
(PDEs) which govern the wave propagation cease to apply at the discontinuities,
the relevant physics, which is represented by the conservation laws
behind those equations, may still hold \citep{Whitham1974}. Thus,
the propagation of hydraulic jumps is expected to be governed not
by the original PDEs but by equivalent integral relationships which
are known as the Rankine-Hugoniot conditions in the gas dynamics.
Such relationships can be obtained for the PDEs of the form 
\[
\partial_{t}P(\vec{u})+\partial_{x}Q(\vec{u})=0
\]
by integrating them across the discontinuity. This type of equation
represents a local conservation law for the quantity $P$ with the
flux $Q$ and the dynamical variables $\vec{u}(x,t).$ Conservation
laws determine the speed at which a jump propagates without using
any information about the flow inside the jump.

For a single fluid layer, there is an infinite number of such conservation
laws \citep[p. 459]{Whitham1974}. For a two-layer system with a free
surface, only six such linearly independent laws exist \citep{Ovsyannikov1979,Montgomery2001,Barros2006}.
For a two-layer system bounded by a rigid lid, an infinite number
of conservation laws is expected \citep{Ovsyannikov1979,Milewski2015}.
However, only three most elementary laws, which describe the conservation
of mass, irrotationality (zero vorticity) and energy are generally
known. No local momentum conservation law appears to be known in this
case. At the same time, the conservation of momentum is known to play
a key role for the hydraulic jumps in single fluid layers. At the
same time, single layer represents the limiting case of a two-layer
system when either the density of the top layer or the depth of the
bottom layer becomes small \citep{Stoker1958}.

The lack of a local momentum conservation law has led to the assumption
that two-layer SW equations are inherently non-conservative \citep{Abgrall2009}
and unable to describe internal hydraulic jumps without additional
closure relations. The latter are usually deduced by dimensional arguments
\citep{Abbott1961} or derived using various semi-empirical integral
models \citep{Baines1995}. For gravity currents, which are created
when a layer of heavier liquid is driven by its weight along the bottom
into a lighter ambient fluid, such a front condition relating the
velocity of propagation with the depth of the layer is the central
result of the celebrated Benjamin's theory \citep{Benjamin1968}.
This hydraulic-type condition and its various empirical extensions
\citep{Klemp1994,Huppert2006} are commonly regarded as essential
for the numerical modeling of gravity currents using the hydrostatic
SW approximation \citep{Ungarish2011,Rotunno2011}.

A number of similar semi-empirical front conditions have been proposed
also for internal bores \citep{Yih1955,Baines1984,Wood1984,Klemp1997}.
Despite the long history of this problem, there is still no comprehensive
theoretical description of internal bores, and new models and front
conditions continue to emerge \citep{Borden2012,Borden2013,Baines2016,Ungarish2018,Fyhn2019}
motivated by the importance such bores play in various geophysical
flows ranging from coastal oceans \citep{Scotti2004} to the inversion
layers in the atmosphere \citep{Christie1992}.

In this paper, we propose a new SW theoretical framework for the analysis
and numerical modeling of interfacial waves containing hydraulic jumps.
In contrast to the previous SW models, no external front conditions
are required in our model. The theory is based on a novel, locally
conservative momentum equation, which is derived from the basic SW
equations. The derived equation contains a free parameter $\alpha,$
which emerges due to the inherent non-uniqueness of the SW momentum
equation for two-layer system of fixed height. The proposed SW framework
is mathematically rigorous and free of phenomenological concepts like
the head loss or energy dissipation in separate layers, which are
fundamental to the conventional control-volume approach. 

The paper is organized as follows. In the next section, we introduce
a two-layer model and derive the SW equations in locally conservative
form for fluids with significantly different as well as nearly equal
densities. Jump conditions for the latter case are derived in section
\ref{sec:Jump}, where we also compare the resulting front speeds
with the predictions of some previous models as well as with experimental
and numerical data. In section \ref{sec:bores}, we consider bores
which can form atop gravity currents and, thus, to connect the latter
to deeper upstream states. The paper is concluded with section \ref{sec:Sum}
where the main results are summarized and the principal differences
between the SW and control-volume approaches are critically discussed.
A more detailed discussion of the key differences between the proposed
SW theory and two alternative approaches is presented in the Appendix.

\section{\label{sec:model}Two-layer SW model}

\begin{figure}
\begin{centering}
\includegraphics[width=0.5\columnwidth]{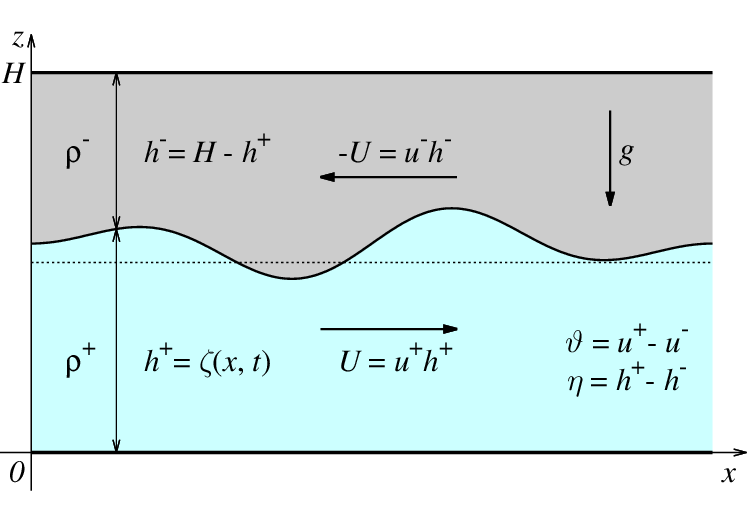}
\par\end{centering}
\caption{\label{fig:sktch1}Sketch of the problem showing a horizontal channel
of constant height $H$ bounded by two parallel solid walls and filled
with two inviscid immiscible fluids with constant densities $\rho^{+}$
and $\rho^{-},$ where $h^{+}=\zeta(x,t)$ and $h^{-}=H-h^{+}$are
the depths of the bottom and top layers, respectively.}
\end{figure}

Consider a horizontal channel of a constant height $H$ which is bounded
by two parallel solid walls and filled with two inviscid immiscible
fluids with constant densities $\rho^{+}$ and $\rho^{-}$ as shown
in Fig. \ref{fig:sktch1}. The fluids are subject to a downward gravity
force with the free fall acceleration $g.$ The interface separating
the fluids at the horizontal position $x$ and the time instant $t$
is located at the height $z=\zeta(x,t).$ The latter is equal to the
depth of the bottom layer $h^{+}=H-h^{-},$ where $h^{-}$ is the
depth of the top layer. The velocity $\vec{u}^{\pm}$ and the pressure
$p^{\pm}$ in each layer are governed by the Euler equation 
\begin{equation}
\partial_{t}\vec{u}+\vec{u}\cdot\vec{\nabla}\vec{u}=-\rho^{-1}\vec{\nabla}p+\vec{g}\label{eq:Euler}
\end{equation}
and the incompressibility constraint $\vec{\nabla}\cdot\vec{u}=0.$
Henceforth, for the sake of brevity, we drop $\pm$ indices wherever
analogous expressions apply to both layers. At the interface $z=\zeta(x,t)$,
we have the continuity of pressure, $[p]\equiv p^{+}-p^{-}=0,$ and
the kinematic condition 
\begin{equation}
w=\frac{d\zeta}{dt}=\zeta_{t}+u\zeta_{x},\label{eq:kin}
\end{equation}
where $u$ and $w$ are the $x$ and $z$ components of velocity,
and the subscripts $t$ and $x$ stand for the corresponding partial
derivatives. Integrating the incompressibility constraint over the
depth of each layer and using Eq. (\ref{eq:kin}), we obtain
\begin{equation}
h_{t}+(h\bar{u})_{x}=0,\label{eq:hpm}
\end{equation}
where the overbar denotes the depth average. Similarly, averaging
the horizontal $(x)$ component of Eq. (\ref{eq:Euler}), we have
\begin{equation}
(h\bar{u})_{t}+(h\overline{u^{2}})_{x}=-\rho^{-1}h\overline{p_{x}}.\label{eq:ub}
\end{equation}
Pressure follows from the integration of the vertical $(z)$ component
of Eq. (\ref{eq:Euler}) as
\begin{equation}
p(x,z,t)=\mathit{\Pi}(x,t)+\rho\int_{\zeta}^{z}(w_{t}+\vec{u}\cdot\vec{\nabla}w-g)dz,\label{eq:prs}
\end{equation}
where the constant of integration $\mathit{\Pi}(x,t)=\left.p^{\pm}(x,z,t)\right|_{z=\zeta}$
defines the distribution of pressure along the interface. Averaging
the $x$-component of the gradient of pressure (\ref{eq:prs}) over
the depth of each layer, after a few rearrangements, we obtain
\begin{equation}
\overline{p_{x}}=\left(\mathit{\Pi}+\rho g\zeta+\rho\overline{(z-z_{0})(w_{t}+\vec{u}\cdot\vec{\nabla}w)}\right)_{x},\label{eq:pxb}
\end{equation}
which defines the RHS of Eq. (\ref{eq:ub}) with $z_{0}=0$ and $z_{0}=H$
for the bottom and top layers, respectively.

In the SW approximation, which is applicable when the characteristic
horizontal length scale $L$ is much larger than the height $H,$
i.e. $H/L=\epsilon\ll1,$ the exact depth-averaged equations above
can be simplified follows. In this case, the incompressibility constraint
implies $w/u=O(\epsilon)$ and Eq. (\ref{eq:pxb}) correspondingly
reduces to 
\begin{equation}
\overline{p_{x}}=(\mathit{\Pi}+\rho g\zeta)_{x}+O(\epsilon^{2}),\label{eq:px0}
\end{equation}
where the leading-order term is purely hydrostatic and $O(\epsilon^{2})$
represents a small dynamical pressure correction due to the vertical
velocity $w.$ Additionally, the flow in each layer is assumed to
be irrotational: $\vec{\omega}=\vec{\nabla}\times\vec{u}=0.$ According
to the inviscid vorticity equation 
\[
\frac{d\vec{\omega}}{dt}=(\vec{\omega}\cdot\vec{\nabla)}\vec{u,}
\]
 this property is preserved by Eq. (\ref{eq:Euler}). In the leading-order
approximation, the irrotationality condition reduces to $\partial_{z}u^{(0)}=0$.
It means that the horizontal velocity can be decomposed as 
\begin{equation}
u=\bar{u}+\tilde{u},\label{eq:ubt}
\end{equation}
where $\tilde{u}$ is the deviation from average which according to
Eq. (\ref{eq:px0}) is $O(\epsilon^{2}).$ Consequently, in the second
term of Eq. (\ref{eq:ub}), we have $\overline{u^{2}}=\bar{u}^{2}+O(\epsilon^{4}).$
Finally, using Eq. (\ref{eq:hpm}) and ignoring $O(\epsilon^{2})$
dynamic pressure correction, Eq. (\ref{eq:ub}) can be written as
\begin{equation}
\rho(\bar{u}_{t}+\tfrac{1}{2}\text{\ensuremath{\bar{u}^{2}}}_{x}+g\zeta_{x})=-\mathit{\Pi}_{x}.\label{eq:upm}
\end{equation}
This and Eq. (\ref{eq:hpm}) constitute the basic set of SW equations
in the leading-order (hydrostatic) approximation.

For completeness, note that the vertical velocity, which is outside
the scope of the present study, can obtained from the incompressibility
constraint and Eq. (\ref{eq:ubt}) as 
\begin{equation}
w(z)=-\int_{z_{0}}^{z}u_{x}dz=-(z-z_{0})\bar{u}_{x}+O(\epsilon^{2}),\label{eq:w}
\end{equation}
where $z_{0}$ is defined as in Eq. (\ref{eq:pxb}) to satisfy the
impermeability conditions $w(0)=w(H)=0.$ Then Eq. (\ref{eq:pxb})
straightforwardly leads to the well-known result \citep{Green1976,Liska1995,Choi1999}
\begin{equation}
h\overline{p_{x}^{(1)}}=-\frac{1}{3}\rho\left(h^{3}(D_{t}\bar{u}_{x}-\text{\ensuremath{\bar{u}_{x}}}^{2}\right)_{x}+O(\epsilon^{4})=\frac{1}{3}\rho\left(h^{2}D_{t}^{2}h\right)_{x}+O(\epsilon^{4}),\label{eq:mcc}
\end{equation}
where $D_{t}\equiv\partial_{t}+\bar{u}\partial_{x}$ and $\bar{u}_{x}=-h^{-1}D_{t}h.$
The latter relation follows from Eq. (\ref{eq:hpm}) and ensures that
kinematic constraint (\ref{eq:kin}) is satisfied by Eq. (\ref{eq:w})
up to $O(\epsilon^{2}).$

On one hand, the wave dispersion caused by the weakly non-hydrostatic
pressure correction (\ref{eq:mcc}) can prevent the development of
discontinuities and enable the formation of solitary waves and permanent-shape
bores (solibores) \citep{Choi1999}. On the other hand, the weakly
non-hydrostatic approximation is limited to relatively shallow waves
and, thus, inapplicable to strong internal bores \citep{Esler2011}.
The latter are the main focus of the present study, where we show
that such bores can be described by the hydrostatic SW approximation
in a self-contained way without invoking externally derived front
conditions.

The system of four SW Eqs. (\ref{eq:upm}) and (\ref{eq:hpm}) contains
five unknowns: $u^{\pm}$, $h^{\pm}$ and $\mathit{\Pi},$ and is
completed by adding the fixed height constraint $\{h\}\equiv h^{+}+h^{-}=H.$
Henceforth, we simplify the notation by omitting the bar over $u$
and use the curly brackets to denote the sum of the enclosed quantities.
Two more unknowns can be eliminated as follows. Firstly, adding the
mass conservation equations for each layer together and using $\{h\}_{t}\equiv0,$
we obtain $\{uh\}=\Phi(t),$ which is the total flow rate. In this
study, the channel is assumed to be laterally closed, which means
$\Phi\equiv0,$ and, thus, $u^{-}h^{-}=-u^{+}h^{+}.$ Secondly, the
pressure gradient $\mathit{\Pi}_{x}$ can be eliminated by subtracting
the two Eqs. (\ref{eq:upm}) one from another. This leaves only two
unknowns, $U\equiv u^{+}h^{+}$ and $h=h^{+},$ and two equations,
which can be written in a locally conservative form as 
\begin{align}
\left(\left\{ \rho/h\right\} U\right)_{t}+\left(\tfrac{1}{2}\left[\rho/h^{2}\right]U^{2}+g\left[\rho\right]h\right)_{x} & =0,\label{eq:u}\\
h_{t}+U_{x} & =0.\label{eq:h}
\end{align}
where the square brackets denote the difference of the enclosed quantities
between the bottom and top layers: $\left[f\right]\equiv f^{+}-f^{-}.$
In this form, both equations can in principle be integrated across
the discontinuity to obtain the corresponding jump conditions. But
this, as it will be shown in the following, is not the only possible
set of locally conservative SW equations.

The applicability of Eqs. (\ref{eq:u}) and (\ref{eq:h}) to strong
bores depends on the conservation of the corresponding quantities
not only in simple one-dimensional flows, which are described explicitly
by these equations, but also in more complex three-dimensional turbulent
flows, which usually occur in strong bores. The conservation of mass
described by Eq. (\ref{eq:h}) in each layer is supposed to hold if
fluids are immiscible. This is assumed in the present study but may
not always be the case \citep{Milewski2015}. The quantity conserved
in Eq. (\ref{eq:u}), which can be written as 
\[
\left\{ \rho/h\right\} U=[\rho u]=\int_{H}\partial_{z}(\rho u)\,dz,
\]
is related to the vorticity. Namely, in each layer separately, we
have $\partial_{z}(\rho u)=\rho\omega,$ where $\omega=\partial_{z}u\equiv0$
is the vorticity in the hydrostatic approximation. On the other hand,
the quantity conserved in Eq. (\ref{eq:u}) is related to $\oint\rho\vec{u}\cdot d\vec{r}=\int_{\mathit{\Gamma}}[\rho u]dx,$
which represents circulation in the $(x,z)$-plane around a small
segment $\mathit{\Gamma}$ of the vortex sheet made by the interface.
It is important to note that the conservation of this quantity is
limited to strictly two-dimensional flows which only advect the vorticity
but do not generate it. This, however, is not the case in three-dimensional
flows in which vorticity can be generated by stretching and twisting
of vortices. It implies that the quantity whose conservation is described
by Eq. (\ref{eq:u}) may not be conserved in strong bores. Also note
that this quantity is not expected to be conserved in single fluid
layers either. The quantity which is expected to be conserved across
hydraulic jumps in single fluid layers with smooth bottom is the momentum
\citep{Kalisch2017}. Analogous quantity can be expected to be conserved
also in two-layer system with flat top and bottom boundaries. 

To obtain momentum equation for the two-layer system we multiply Eq.
(\ref{eq:upm}) for each layer with $h^{\pm}$ and add both equations
together. Using Eq. (\ref{eq:h}) along with the fixed-height condition,
after a few rearrangements, we have 
\begin{equation}
[\rho]U_{t}+\left(\{\rho/h\}U^{2}+\tfrac{1}{2}g[\rho h^{2}]+H\mathit{\Pi}\right)_{x}=0,\label{eq:Up}
\end{equation}
where $[\rho]U\equiv\{\rho uh\}$ is the momentum density. In this
form, the momentum conservation equation is non-local because it contains
not only the dynamical variables $U$ and $h$ but also the interfacial
pressure $\mathit{\Pi}$. The latter can be eliminated from Eq. (\ref{eq:Up})
in two alternative ways. First, if we follow the same steps as in
deriving Eq. (\ref{eq:Up}), but before adding the two equations together
divide them by $\rho^{\pm},$ we obtain 
\[
\mathit{\Pi}_{x}=-\{h/\rho\}^{-1}\left(\{h^{-1}\}U^{2}+gHh\right)_{x}.
\]
 Although substituting this expression into Eq. (\ref{eq:Up}) we
can eliminate $\mathit{\Pi}_{x},$ it does not render the resulting
equation locally conservative. The problem is the non-local dependence
of the pressure $\mathit{\Pi=\int\mathit{\Pi}_{x}}\thinspace dx$
on the dynamical variables $U$ and $h.$ But it does not mean that
Eq. (\ref{eq:Up}) is inherently non-local as it is commonly believed.
The alternative approach which allows us to cast this equation into
locally conservative form, is to take $\mathit{\Pi}_{x}$ directly
from Eq. (\ref{eq:upm}). But here we are faced with a dilemma as
$\mathit{\Pi}_{x}$ can be taken either from the equation for the
top or bottom layer \citep{Ostapenko1999}. Note that requiring Eq.
(\ref{eq:upm}) to yield the same $\mathit{\Pi}_{x}$ for both layers
leads back to Eq. (\ref{eq:u}), which, as discussed above, describes
the conservation of circulation. Therefore, the two expressions of
$\mathit{\Pi}_{x}$ following from Eq. (\ref{eq:upm}) are equivalent
only if Eq. (\ref{eq:u}) is satisfied. But this, as shown in the
following, is not in general possible. 

We resolve this dilemma by taking a linear combination of the pressure
gradients defined by Eqs. (\ref{eq:upm}) for each layer with weight
coefficients $(1\pm\alpha)/2,$ where $\alpha$ is an arbitrary constant
defining the contribution of each layer to $\mathit{\Pi}_{x}.$ This
results in 
\begin{equation}
\mathit{\Pi}_{x}=-\tfrac{1}{2}\left(\left(\left[\rho/h\right]U\right)_{t}+\left(\tfrac{1}{2}\left\{ \rho/h^{2}\right\} U^{2}+g\left\{ \rho\right\} h\right)_{x}\right)-\tfrac{1}{2}\alpha\mathit{\Lambda},\label{eq:Px}
\end{equation}
where 
\begin{equation}
\mathit{\Lambda}=\left(\left\{ \rho/h\right\} U\right)_{t}+\left(\tfrac{1}{2}\left[\rho/h^{2}\right]U^{2}+g\left[\rho\right]h\right)_{x}\label{eq:Lambda}
\end{equation}
is the LHS of Eq. (\ref{eq:u}). As seen from the definition of weight
coefficients above$,$ $\mathit{\Lambda}$ represents the difference
of $\mathit{\Pi}_{x}$ between the values defined by Eqs. (\ref{eq:upm})
for the bottom and top layers. Note that with $\alpha=1$, $\mathit{\Pi}_{x}$
is determined solely by the top layer, whereas the opposite is the
case with $\alpha=-1.$ In general, we can also have $|\alpha|>1$
as one weight coefficient may be negative while the other is greater
than unity. If Eq. (\ref{eq:u}) is satisfied, i.e. $\mathit{\Lambda}=0$,
the last term in Eq. (\ref{eq:Px}) with $\alpha$ vanishes. Then
substituting $\mathit{\Pi}_{x}$ from Eq. (\ref{eq:Px}) into Eq.
(\ref{eq:Up}), we obtain
\begin{equation}
\left([\rho-\tfrac{1}{2}H\rho/h]U\right)_{t}+\left(\{\rho/h-\tfrac{1}{4}H\rho/h^{2}\}U^{2}+\tfrac{1}{4}g[\rho]\{h^{2}\}\right)_{x}=0,\label{eq:U}
\end{equation}
which is the two-layer momentum equation (\ref{eq:Up}) written in
locally conservative form. Note that it is not the momentum $[\rho]U$
but rather the pseudo-momentum $[\rho-\tfrac{1}{2}H\rho/h]U,$ called
the impulse by \citet{Benjamin1984}, which emerges as a conserved
quantity in this equation. It reflects the inconspicuous fact that
it is the pseudo-momentum rather than the momentum which is actually
conserved in the laterally closed two-layer system bounded by a rigid
lid \citep{Benjamin1986,Camassa2012}. Equation (\ref{eq:U}) is
equivalent to Eq. (\ref{eq:u}) and can be reduced to the latter by
using Eq. (\ref{eq:h}) provided that both $U$ and $h$ are differentiable
at least once. This is obviously not so at the points where $U$ and
$h$ are discontinuous. In this case, Eqs. (\ref{eq:u}) and (\ref{eq:U})
cannot in general be satisfied simultaneously. It means that we cannot
assume $\mathit{\Lambda}=0$ when substituting $\mathit{\Pi}_{x}$
from Eq. (\ref{eq:Px}) into Eq. (\ref{eq:Up}). Thus, the term $-\tfrac{1}{2}\alpha H\mathit{\Lambda}$
has to be retained in Eq. (\ref{eq:U}) for this equation to be applicable
also to discontinuous solutions. Since Eq. (\ref{eq:Lambda}) defining
$\mathit{\Lambda}$ is locally conservative, so is also the momentum
equation containing the extra term with $\alpha\mathit{\Lambda}:$
\begin{multline}
\left(\left([\rho-\tfrac{1}{2}H\rho/h]-\tfrac{1}{2}\alpha H\left\{ \rho/h\right\} \right)U\right)_{t}+\\
\left(\left(\{\rho/h-\tfrac{1}{4}H\rho/h^{2}\}-\tfrac{1}{4}\alpha H\left[\rho/h^{2}\right]\right)U^{2}+\tfrac{1}{2}g[\rho]\{\tfrac{1}{2}h^{2}-\alpha Hh\}\right)_{x}=0.\label{eq:U-gen}
\end{multline}
Subsequently, this equation will be referred to as the generalized
momentum equation.

Since $\alpha$ is a dimensionless constant, it can depend only the
ratio of densities, which is the sole dimensionless parameter in this
problem. As $\mathit{\Pi_{x}}$ is expected to vanish when the top
layer density $\rho^{-}$ becomes small and, thus, the two-layer system
reduces to single layer, Eq. (\ref{eq:upm}) suggests that, in this
limit, $\mathit{\Pi_{x}}$ has to be determined solely by the top
layer. As discussed above, this corresponds to $\alpha\rightarrow1$.
In the opposite limit of a small density difference, one can expect
$\alpha\rightarrow0,$ which corresponds to both layers affecting
$\mathit{\Pi_{x}}$ with equal weight coefficients.

In the following, the propagation velocities of internal bores resulting
from the mass conservation equation (\ref{eq:h}) and the generalized
momentum equation (\ref{eq:U-gen}) with various $\alpha$ will be
considered and compared with the available experimental and numerical
results.

To determine which of several possible solutions are physical, we
will need also an energy equation. Multiplying Eq. (\ref{eq:upm})
for each layer with $U$ and using Eq. (\ref{eq:h}), we obtain 
\begin{equation}
\rho\left(U^{2}/h\pm gh^{2}\right)_{t}+\left(\left(\rho\left(U^{2}/h^{2}\pm2gh\right)+2\mathit{\Pi}\right)U\right)_{x}=-2\mathit{\Pi}h_{t}.\label{eq:Ep}
\end{equation}
where $h$ and $U$ stand for $h^{\pm}$ and $U^{\pm},$ respectively,
and the plus and minus signs correspond as usual to the bottom and
top layers. These are two intermediate equations which govern the
energy of separate layers. As seen, the RHS term, which describes
the energy exchange between the layers, makes these equations non-conservative.
Therefore, the energy is not conserved in each layer separately unless
the RHS term vanishes. This is usually taken for granted in the control-volume
approach, where the flow in the hydraulic jump is assumed to be stationary
in the co-moving frame of reference. This, however, is not likely
to be the case for the bores which are either turbulent or undular.
It is important to note that in the hydrostatic SW approximation,
a velocity distribution that is stationary cannot be continuous. Therefore,
the conservation of energy in separate fluid layers is in general
mathematically incompatible with the hydrostatic SW approximation.
There is, however, one exception corresponding to the so-called solibores,
which will be considered later.

Owing to the fixed height constraint $\{h\}_{t}=0,$ the RHS terms
in Eqs. (\ref{eq:Ep}) cancel out when both equations are added together.
As a result, we have

\begin{equation}
\left(\{\rho/h\}U^{2}+\tfrac{1}{4}g[\rho][h^{2}]\right){}_{t}+\left(\left([\rho/h^{2}])U^{2}+g[\rho][h]\right)U\right)_{x}=0.\label{eq:E}
\end{equation}
This locally conservative two-layer energy equation is used in the
following to discriminate unphysical solutions.

The local mass, circulation, momentum and energy conservation laws
which are defined respectively by Eqs. (\ref{eq:h}), (\ref{eq:U})
and (\ref{eq:E}), can be integrated across discontinuities to obtain
jump conditions analogous to the Rankine-Hugoniot relations and the
Lax entropy constraint in the gas dynamics. Since Eqs. (\ref{eq:u}),
(\ref{eq:U}) and (\ref{eq:E}) are mutually equivalent and can be
transformed one into another using Eq. (\ref{eq:h}) only if $h$
and $U$ are continuous, the jump conditions resulting from these
equations cannot in general be satisfied simultaneously. As the problem
is governed by two equations, only two corresponding jump conditions
can be satisfied. The choice of two quantities which can be conserved
across the jump is not obvious and depends on additional physical
arguments. Namely, it depends on the effects, such as the viscous
dissipation, three-dimensional vorticity generation and mixing (entrainment),
which are ignored in the SW approximation but can become relevant
in hydraulic jumps. 

If the SW approximation breaks down in a relatively narrow region,
then the complex phenomena taking place in that region can be taken
into account by applying the relevant conservation laws and treating
the region as a discontinuity \citep{Whitham1974}. As already noted,
since the hydrostatic SW model is a long-wave approximation, the variation
of flow over the horizontal length scale comparable to the layer depth
or shorter appears as a discontinuity.

In the following, we assume the density difference to be small as
it is often the case in reality. Then, according to the Boussinesq
approximation, the density difference can be neglected for the inertia
but not for the gravity of fluids.  We slightly extend this approximation
by neglecting the deviation of the density form its average value.
The latter is subsequently used as a characteristic value instead
of the density of one of the layers. Then Eq. (\ref{eq:U}) reduces
to 
\begin{equation}
\left([h][u]\right){}_{t}+\tfrac{1}{4}\left((H-3[h]^{2}/H)[u]^{2}+2g[h]^{2}[\rho]/\{\rho\}\right){}_{x}=0.\label{eq:uh}
\end{equation}
The problem can be simplified further by using the total height $H$
and the characteristic gravity wave speed $C=\sqrt{2Hg[\rho]/\{\rho\}}$
as a vertical length scale and a velocity scale, respectively. We
use $L$ as a horizontal length scale and $L/C$ as a time scale.
Then the basic momentum equation (\ref{eq:uh}) and the total energy
equation (\ref{eq:E}) can be written in dimensionless form as
\begin{align}
(\eta\vartheta)_{t}+\tfrac{1}{4}(\eta^{2}+\vartheta^{2}-3\eta^{2}\vartheta^{2})_{x} & =0,\label{eq:mnt}\\
(\eta^{2}+\vartheta^{2}-\eta^{2}\vartheta^{2})_{t}+(\eta\vartheta(1-\eta^{2})(1-\vartheta^{2}))_{x} & =0,\label{eq:nrg}
\end{align}
where $\eta=[h]$ and $\vartheta=[u]$ are the depth and velocity
differentials between the bottom and top layers. These two quantities
emerge as natural variables for this problem. Subsequently, the former
is referred to as the interface height and the latter as the shear
(or baroclinic) velocity. In the new variables and the Boussinesq
approximation, Eqs. (\ref{eq:u}) and (\ref{eq:h}), which describe
the conservation of circulation and mass, respectively, take a remarkably
symmetric form \citep{Milewski2015}
\begin{align}
\vartheta_{t}+\tfrac{1}{2}(\eta(1-\vartheta^{2}))_{x} & =0,\label{eq:crc}\\
\eta_{t}+\tfrac{1}{2}(\vartheta(1-\eta^{2}))_{x} & =0.\label{eq:vlm}
\end{align}
Correspondingly, the generalized momentum equation (\ref{eq:U-gen})
reads as

\begin{equation}
((\eta+\alpha)\vartheta)_{t}+\tfrac{1}{4}(\eta^{2}+\vartheta^{2}-3\eta^{2}\vartheta^{2}+2\alpha\eta(1-\vartheta^{2}))_{x}=0.\label{eq:gen}
\end{equation}
Note that this equation represents a linear combination of Eqs. (\ref{eq:mnt})
and (\ref{eq:crc}) in which the latter is multiplied with $\alpha.$
Therefore, Eq. (\ref{eq:gen}) reduces to the basic momentum equation
(\ref{eq:mnt}) when $\alpha=0,$ and to the circulation conservation
equation (\ref{eq:crc}) when |$\alpha|\rightarrow\infty.$

Note that owing to the equivalence of various local conservation laws
for continuous solutions, Eqs. (\ref{eq:mnt}) and (\ref{eq:nrg})
can be derived directly from Eqs. (\ref{eq:crc}) and (\ref{eq:vlm}).
Moreover, an infinite sequence of hyperbolic conservation laws can
be constructed starting from the basic equations (\ref{eq:crc},\ref{eq:vlm})
\citep{Milewski2015}. The basic equations can also be written in
the canonical form 
\begin{equation}
R_{t}^{\pm}+\lambda^{\pm}R_{x}^{\pm}=0,\label{eq:canon}
\end{equation}
where $R^{\pm}=-\eta\vartheta\pm\sqrt{(1-\eta^{2})(1-\vartheta^{2})}$
are the Riemann invariants and 
\begin{equation}
\lambda^{\pm}=\frac{3}{4}R^{\pm}+\frac{1}{4}R^{\mp}\label{eq:Cpm}
\end{equation}
are the associated characteristic velocities \citep{Long1956,Cavanie1969,Ovsyannikov1979,Sandstrom1993,Baines1995,Chumakova2009}.

Since the interface is confined between the top and bottom boundaries,
which corresponds to $\eta^{2}\le1,$ the characteristic velocities
(\ref{eq:Cpm}) are real and, thus, the equations are of hyperbolic
type if $\vartheta^{2}\le1.$ The solutions that do not satisfy the
latter constraint are subject to a long-wave shear instability and,
thus, physically infeasible \citep{Milewski2004,Esler2011}.

\section{\label{sec:Jump}Jump conditions}

\begin{figure}
\begin{centering}
\includegraphics[width=0.5\columnwidth]{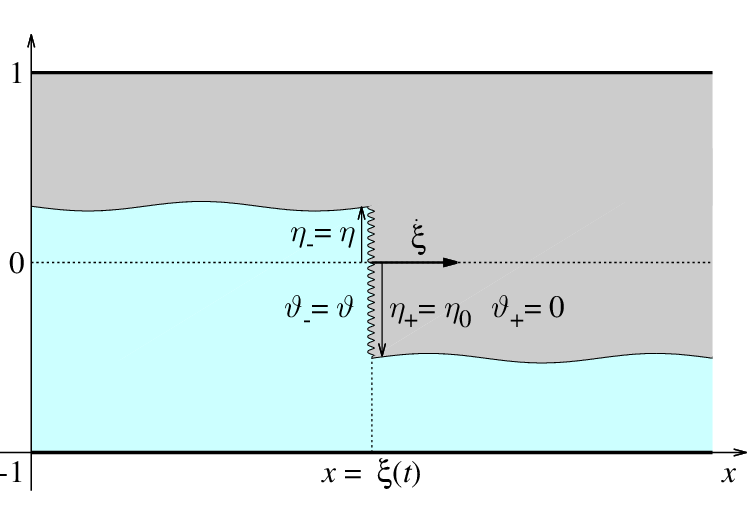}
\par\end{centering}
\caption{\label{fig:sktch2}A jump with the upstream interface height $\eta_{-}=\eta$
and the shear velocity $\vartheta_{-}=\vartheta$ propagating at the
speed $\dot{\xi}$ into a still fluid ahead $(\vartheta_{+}=0)$ with
the interface located at the height $\eta_{+}=\eta_{0}.$}
\end{figure}
Consider a hydraulic jump which occurs over a length scale comparable
to the layer depth and thus appears in the SW approximation as a discontinuity
in $\eta$ and $\vartheta$ at the point $x=\xi(t)$ across which
the respective variables jump by $\left\llbracket \eta\right\rrbracket \equiv\eta_{+}-\eta_{-}$
and $\left\llbracket \vartheta\right\rrbracket \equiv\vartheta_{+}-\vartheta_{-}.$
Here the plus and minus subscripts denote the corresponding quantities
at the front and behind of the jump. The double-square brackets stand
for the differential of the enclosed quantity across the jump. Integrating
Eqs. (\ref{eq:vlm}) and (\ref{eq:gen}) across the jump, which is
equivalent to substituting spatial derivative $f_{x}$ with $\left\llbracket f\right\rrbracket $
and time derivative $f_{t}$ with $-\dot{\xi}\left\llbracket f\right\rrbracket $
\citep{Whitham1974}, the jump propagation velocity can be expressed,
respectively, as 
\begin{eqnarray}
\dot{\xi} & = & \frac{1}{2}\frac{\left\llbracket \vartheta(1-\eta^{2})\right\rrbracket }{\left\llbracket \eta\right\rrbracket },\label{eq:jmp1}\\
\dot{\xi} & = & \frac{1}{4}\frac{\left\llbracket \eta^{2}+\vartheta^{2}-3\eta^{2}\vartheta^{2}+2\alpha\eta(1-\vartheta^{2})\right\rrbracket }{\left\llbracket (\alpha+\eta)\vartheta\right\rrbracket }.\label{eq:jmp2}
\end{eqnarray}
 As for single layer, the jump conditions consist of two equations
and contain five unknowns: $\eta_{\pm},$ $\vartheta_{\pm}$ and $\dot{\xi}.$
Consequently, two unknown parameters can be determined when the other
three are specified. Since the jump conditions are non-linear, multiple
solutions are possible. Some of these solutions may be unphysical.
Feasible solutions are selected by an additional constraint which
follows from the energy equation (\ref{eq:nrg}). Integrating this
equation as described above, we obtain the following difference of
energy fluxes across the jump 
\begin{equation}
\left\llbracket \eta\vartheta(1-\eta^{2})(1-\vartheta^{2})-\dot{\xi}\left(\eta^{2}+\vartheta^{2}-\eta^{2}\vartheta^{2}\right)\right\rrbracket =\dot{\varepsilon}\le0.\label{eq:deps}
\end{equation}
This quantity cannot be positive because there is no physical mechanism
which could generate energy in the jump. Energy can be either dissipated
or dispersed by the short non-hydrostatic waves excited by the jump
\citep{Ali2010}. 

\begin{figure}
\centering{}\includegraphics[width=0.3\columnwidth]{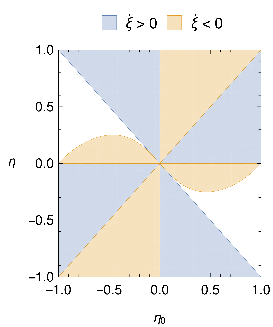}\put(-125,155){(\textit{a})}
\includegraphics[width=0.3\columnwidth]{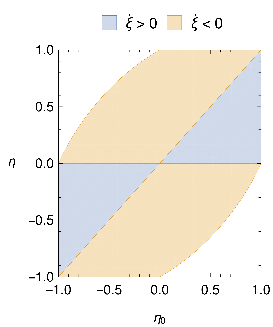}\put(-125,155){(\textit{b})}\\
\includegraphics[width=0.3\columnwidth]{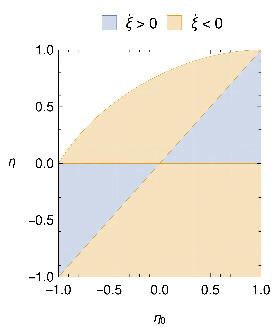}\put(-125,155){(\textit{c})}
\includegraphics[width=0.3\columnwidth]{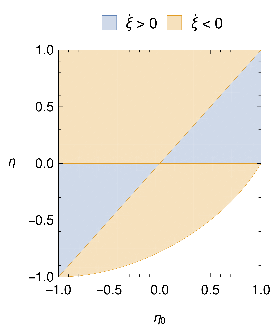}\put(-125,155){(\textit{d})}\caption{\label{fig:h0h}The downstream $(\eta_{0})$ and upstream $(\eta)$
heights of bores permitted by the hyperbolicity constraint $\vartheta^{2}\le1$
for $\alpha=0$ (a), $\infty$ (b), $-1$ (c), $1$ (d). The downstream
$(\dot{\xi}>0)$ and upstream $(\dot{\xi}<0)$ directions of propagation
are defined by the energy constraint $\dot{\varepsilon}\le0.$}
\end{figure}

Next, let us apply general jump conditions (\ref{eq:jmp1},\ref{eq:jmp2})
to a bore with the upstream interface height $\eta_{-}=\eta$ which
propagates into a quiescent fluid $(\vartheta_{+}=0)$ with the interface
located at the height $\eta_{+}=\eta_{0},$ as shown in Fig. \ref{fig:sktch2}.
After a few rearrangements, the upstream shear velocity $\vartheta_{-}=\vartheta$
and the propagation speed can expressed, respectively, as

\begin{align}
\vartheta^{\pm} & =\pm\frac{(\eta_{0}-\eta)(\eta_{0}+\eta+2\alpha)^{1/2}}{((1-\eta^{2})(\eta_{0}-\eta)+2(\eta+\alpha)(1-\eta_{0}\eta))^{1/2}},\label{eq:v2}\\
\dot{\xi}^{\pm} & =-\vartheta^{\pm}\frac{1-\eta^{2}}{2(\eta_{0}-\eta)},\label{eq:xid}
\end{align}
where the plus and minus signs refer to the opposite directions of
propagation, i.e., $\dot{\xi}^{+}=-\dot{\xi}^{-},$ which are both
permitted by the mass and momentum balance conditions. Because the
energy balance (\ref{eq:deps}) changes sign with the direction of
propagation, only one direction is usually permitted for the bore
of given height. The possible downstream $(\eta_{0})$ and upstream
$(\eta)$ heights of bores permitted by the hyperbolicity constraint
$(\vartheta^{2}\le1)$ and their direction of propagation are shown
in Fig. \ref{fig:h0h} for various $\alpha.$ As discussed before,
$\alpha=0$ corresponds to the basic momentum equation (\ref{eq:mnt}),
in which both layers contribute equally to the pressure drop across
the jump. $\alpha=1$ and $\alpha=-1$ correspond to the pressure
drops determined by the top and bottom layer, respectively. $\alpha\rightarrow\infty$
corresponds to the circulation conservation law (\ref{eq:crc}), which
is based on the assumption that the pressure drops across the discontinuity
in both layers are equal. The downstream and upstream heights which
satisfy the hyperbolicity constraint depend on $\text{\ensuremath{\alpha}.}$
For each such combination of heights, bore can propagate either downstream
$(\dot{\xi}>0)$ or upstream $(\dot{\xi}<0)$ depending on the energy
constraint (\ref{eq:deps}). As seen in Fig. \ref{fig:h0h}, the respective
regions in the $(\eta_{0},\eta)$ plane for $\alpha=0$ and $\alpha=\infty$
are centrally symmetric, whereas for $\alpha=\pm1$ they are centrally
reflected images of each other. The last two values exclude the bores
with $\eta_{0}\rightarrow\pm1,$ which correspond to deep and shallow
downstream states, respectively.

In all four cases, the direction of propagation can be seen in Fig.
\ref{fig:h0h} to reverse along two lines: $\eta=\eta_{0}$ and $\eta=0.$
The first (diagonal) line corresponds to infinitesimal-amplitude waves.
In this limit, the propagation speed (\ref{eq:xid}) becomes equal
to the characteristic velocity (\ref{eq:Cpm}). The second (horizontal
middle) line corresponds to the bores with the upstream interface
located at the channel mid-height. These bores are exceptional. First,
in contrast to all other bores, they conserve the energy and, thus,
can propagate in either direction. Second, their velocity of propagation
is independent of their height and equal to $\pm\frac{1}{2}$ for
all $\alpha$. It means that these bores conserve also the circulation.
These are exactly the properties of the so-called solibores (see Eqs.
(3.33) and (3.34) in \citep{Esler2011}) which appear as permanent-shape
solutions in the weakly non-hydrostatic approximation described by
Eq. (\ref{eq:mcc}).

\begin{figure}
\centering{}\includegraphics[width=0.5\columnwidth]{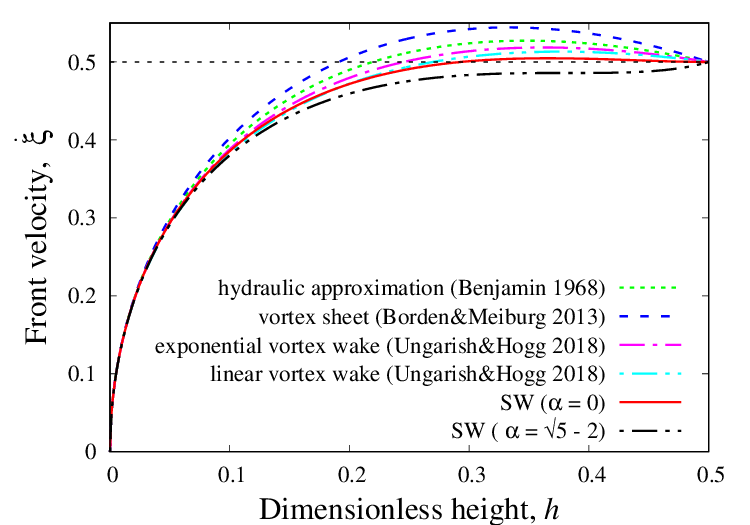} \caption{\label{fig:gc}The front velocity of gravity current $\dot{\xi}$
versus the dimensionless front height $h=\frac{1}{2}(1+\eta):$ comparison
of the SW result for $\alpha=0$ and $\alpha=\sqrt{5}-2$ with the
classical hydraulic approximation due to \citet{Benjamin1968}, the
vortex sheet model of \citet{Borden2013} and the vortex-wake model
of \citet{Ungarish2018}.}
\end{figure}

Gravity currents correspond to the limiting case of bores which propagate
along the bottom $(\eta_{0}=-1).$ As seen in Fig. \ref{fig:h0h},
the depth of gravity currents is limited to the channel mid-height
$(\eta\le0)$ for all considered values of $\alpha$ except $\alpha=1.$
For a gravity current which propagates downstream, Eqs. (\ref{eq:v2})
and (\ref{eq:xid}) yield 
\begin{eqnarray}
\vartheta & = & \left(\frac{(1+\eta)(1-\eta-2\alpha)}{1-\eta^{2}-2(\alpha+\eta)}\right)^{1/2},\label{eq:vgc}\\
\dot{\xi} & = & \frac{1}{2}(1-\eta)\vartheta.\label{eq:xidgc}
\end{eqnarray}
The propagation velocity (\ref{eq:xidgc}) can be written in terms
of the traditional front height $h=(1+\eta)/2$ as 
\begin{equation}
\dot{\xi}=(1-h)\sqrt{\frac{2h(1-\alpha-h)}{1-\alpha-2h^{2}}}.\label{eq:xidh}
\end{equation}
 As seen in Fig. \ref{fig:gc} for $\alpha=0,$ this SW front velocity
is generally slightly lower than that resulting from the well-known
Benjamin's formula $\dot{\xi}=\sqrt{\frac{h(1-h)(2-h)}{1+h}}$ \citep{Benjamin1968},
whereas the vortex-sheet model of \citet{Borden2013} yields a somewhat
higher front velocity $\dot{\xi}=(1-h)\sqrt{2h}.$ It is noteworthy
that the same propagation velocity results also from the SW circulation
conservation equation (\ref{eq:crc}) which corresponds to $|\alpha|\rightarrow\infty$
in Eq. (\ref{eq:xidh}).

It turns out that also Benjamin's formula follows from Eq. (\ref{eq:xidh})
with $\alpha=-1,$ which corresponds to the pressure along the interface
determined solely by the bottom layer. In the control-volume approach,
this is interpreted as the conservation of energy in the bottom layer.
Such an interpretation is based on the assumption that the flow in
the hydraulic jump is stationary. As argued above in relation to Eq.
(\ref{eq:Ep}), such an assumption is not compatible with the hydrostatic
SW approximation. It is also interesting to note that Eq. (\ref{eq:xidh})
reproduces the general vortex-sheet formula derived by \citet{Ungarish2018}
when the ratio of the so-called head losses in the top and bottom
layer is substituted with $\frac{\alpha+1}{\alpha-1}.$ Implications
of this rather non-obvious mathematical equivalence will be discussed
in the conclusion.

We include in Fig. \ref{fig:gc} also the recent results of \citet{Ungarish2018}
for gravity currents obtained using the vortex-wake model in which
a shear layer of finite-thickness is assumed instead of sharp interface.
The assumption of a diffuse interface takes this model outside the
scope of the SW approximation. All models can be seen to yield the
same velocity for thin layers $(h\rightarrow0):$ $\dot{\xi}/\sqrt{h}\rightarrow\sqrt{2},$
which is the classical result due to von Kármán \citep{Huppert2006},
as well as for the gravity currents spanning the lower half of the
channel $(h=1/2):$ $\dot{\xi}/\sqrt{h}=1/\sqrt{2}.$ For intermediate
heights, the SW model with $\alpha=0$ produces generally lower front
velocities than the previous models.

Based on the experimental observations, it has been suggested by \citet{Rottman1983}
and \citet{Marino2005} that for shallow gravity currents, the normalized
front velocity $\dot{\xi}/\sqrt{h}$ may be closer to $1$ rather
than $\sqrt{2}.$ Numerical results indicate that this discrepancy
may be due to turbulent interfacial drag \citep{Klemp1994} or viscosity
\citep{Haertel2000}. The latter can have a significant effect even
on relatively deep gravity currents up to Reynolds numbers of $O(10^{4})$,
which are typical for laboratory experiments. Alternatively, it may
be due to the uncertainty in the depth of turbulent gravity currents.
As shown in the next section, shallow gravity currents can be connected
to a range of deeper upstream states. Taking the upstream depth as
the front height results in a lower-than-expected normalized front
velocity.

\begin{figure}
\centering{}\includegraphics[width=0.5\columnwidth]{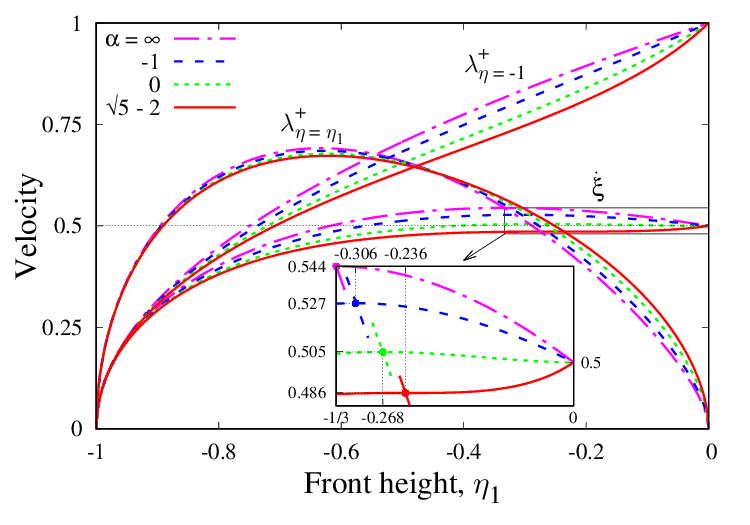}\put(-235,155){(\textit{a})}\includegraphics[width=0.5\columnwidth]{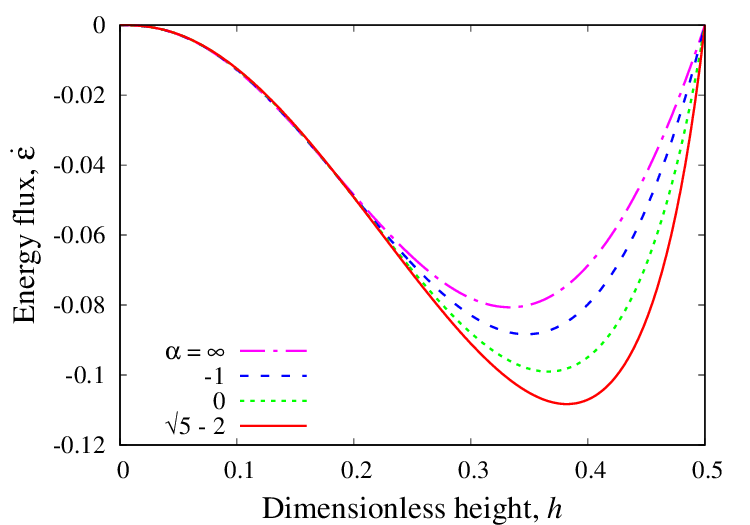}\put(-235,155){(\textit{b})}\caption{\label{fig:hcrt}The characteristic downstream wave speed $\lambda^{+}$
at the top $(z=h)$ and bottom $(z=0)$ of the front and the front
velocity $\dot{\xi}$ versus its height for $\alpha=\infty,-1,0,\sqrt{5}-2$
(a) and the corresponding energy dissipation (b).}
\end{figure}

\citet{Klemp1994} argue that causality does not permit gravity current
to move faster than the characteristic wave velocity (\ref{eq:Cpm})
\[
\lambda^{+}=-\eta\vartheta+\frac{1}{2}\sqrt{(1-\eta^{2})(1-\vartheta^{2})}.
\]
If so, the gravity current height $h=(1+\eta)/2$ cannot exceed
\begin{equation}
h_{c}=\begin{cases}
(\sqrt{3}-1)/2, & \alpha=0,\\
2\sin(\pi/18), & \alpha=-1,\\
1/3, & \alpha\rightarrow\infty,
\end{cases}\label{eq:hc}
\end{equation}
where the last last two values correspond to the front conditions
of \citet{Benjamin1968} and \citet{Borden2013}. It has to be noted,
however, that the front moving at a supercritical speed (faster than
the disturbances behind it) does not violate causality as long as
there are faster moving disturbance ahead of it. This is indeed the
case for the disturbances at the bottom of gravity current (see Fig.
\ref{fig:hcrt}a).

There are two more noteworthy coincidences which occur at the critical
height. First, as seen in the inset of Fig. \ref{fig:hcrt}, the point
at which the characteristic velocity $\left.\lambda_{+}\right|_{z=h}$
drops below the front speed $\dot{\xi}$ for the corresponding height
at given $\alpha$ coincides with the maximum of $\dot{\xi}.$ \citet{Baines2016}
following \citet{Benjamin1968} argue that the presence of such a
maximum implies that gravity currents with $h>h_{c}$ are unstable
and, thus, physically impossible. Namely, if $h\ge h_{c}$ and correspondingly
$\frac{d\dot{\xi}}{dh}<0,$ then a virtual perturbation that reduces
the front height $h$ would increase the front speed $\dot{\xi.}$
By the mass conservation, this would further reduce the front height
thus enhancing the initial perturbation. As a result, the gravity
current would collapse to a stable subcritical height $h\le h_{c}.$
This is a physical mechanism which can limit the height of gravity
current to the critical value (\ref{eq:hc}). Alternatively, by the
same arguments, the instability can result in the increase of the
front height up the channel mid-height. According to Fig. \ref{fig:h0h},
this is the maximal height gravity current can have for all considered
values of $\alpha$ except $\alpha=1.$

The second coincidence, pointed out already by \citet{Benjamin1968},
concerns the energy dissipation rate which also attains a maximum
at exactly the same critical height. The underlying mechanism and
consequences of these two non-obvious coincidences will be elucidated
in the next section, where we consider the possible upstream states
to which gravity current can be connected via an intermediate bore.

It has to be noted that the coincidence of the maximal front propagation
velocity with the characteristic wave speed pointed out above is limited
to 
\begin{equation}
\alpha<\alpha_{c}=\sqrt{5}-2\approx0.236.\label{eq:acrt}
\end{equation}
 The same applies also to the occurrence of maximal dissipation rate
at the critical height. For $\alpha>\alpha_{c},$ the intersection
point of the propagation and characteristic velocities switches over
to the minimum of $\dot{\xi}$ which emerges at $\eta=-\alpha<0$
and moves towards the maximum as $\alpha$ rises above zero. At $\alpha=\alpha_{c},$
the minimum and maximum of $\dot{\xi}$ merge forming a stationary
inflection point at the critical interface height $\eta_{c}=-\alpha_{c},$
where $\partial_{\eta}\dot{\xi}=\partial_{\eta}^{2}\dot{\xi}=0.$
The front height and the propagation speed at this point are $h_{c}=(1-\alpha_{c})/2\approx0.382$
and $\dot{\xi_{c}}=\alpha_{c}^{1/2}\approx0.486$ (see Fig. \ref{fig:hcrt}).
At $\alpha>\alpha_{c},$ the inflection point vanishes as two local
extrema of $\dot{\xi}$ re-emerge. At $\alpha=\frac{1}{2},$ the minimum
of $\dot{\xi}$ moves back to $h=\frac{1}{2}.$ At this point, the
maximum of energy dissipation rate switches from the maximum to minimum
of $\dot{\xi}.$

\begin{figure}
\centering{}\includegraphics[width=0.5\columnwidth]{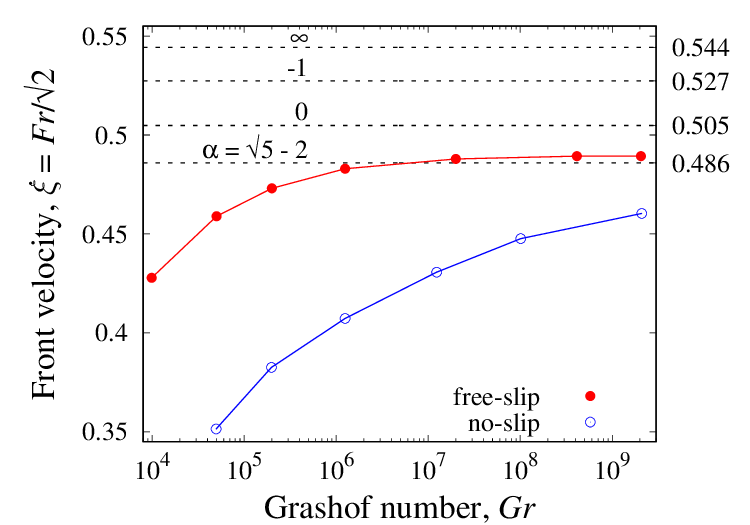} \caption{\label{fig:lckx}Comparison of critical gravity current speeds for
various $\alpha$ with the numerical results of \citet{Haertel2000}
for gravity currents generated by the lock exchange with free-slip
and no-slip boundary conditions. The conversion factor of $1/\sqrt{2}$
is due to the channel half-height used as the length scale in the
definition of Froude number $\mathit{Fr}$ by \citet{Haertel2000};
Grashof number defines the magnitude of the driving force and thus
the characteristic flow velocity with the Reynolds number $\mathit{Re}\sim\sqrt{\mathit{Gr}}.$}
\end{figure}

Thus, $\alpha_{c}$ represents an exceptional point at which the gravity
current speed becomes a monotonically increasing function of its depth.
It is interesting to note that the propagation velocity at this point
agrees surprisingly well with the highly-accurate numerical results
of \citet{Haertel2000} for the gravity currents generated by the
full lock exchange with free-slip boundary conditions (see Fig. \ref{fig:lckx}).
With real no-slip boundary conditions, a much higher Reynolds number
seems to be required to attain this inviscid limit. As shown below,
$\alpha_{c}$ produces a remarkably good agreement with numerical
results not only for gravity currents but also for a wide range of
bores in Boussinesq fluids.

Let us now turn to bores and compare their propagation velocities
resulting from the SW theory with the predictions of some previous
models as well as with the available experimental and numerical data.
For comparison, we choose the semi-empirical model of Klemp, Rotunno
and Skamarock (KRS) \citep{Klemp1997}, the vortex-sheet model of
Borden and Meiburg (BM) \citep{Borden2013} and the vortex-wake model
of Ungarish and Hogg (UH) \citep{Ungarish2018}. The KRS model is
known to achieve a better agreement with the experimental results
by assuming that energy is dissipated only in the top layer, which
shrinks as the bore advances. The BM model is based on the 2D vorticity
equation which is applied in the integral form to bores in Boussinesq
fluids. As for the gravity currents, the BM model yields exactly
the same front speed as the SW circulation conservation law (\ref{eq:crc}):
$\dot{\xi}=\frac{1}{2}(1-\eta^{2})/(1-\eta_{0}\eta)^{1/2}.$ In the
UH model, the conservation of both the circulation and momentum is
effectively imposed in addition to that of the mass. This is not possible
using only the height averaged quantities in each layer, as in the
SW approximation, and requires a non-uniform vertical velocity distribution.
The latter is introduced by replacing the sharp interface with a single-parameter
shear layer. The form of this layer is not uniquely defined and affects
the results as it may be seen in Fig. \ref{fig:gc}.

\begin{figure}
\centering{}\includegraphics[width=0.5\columnwidth]{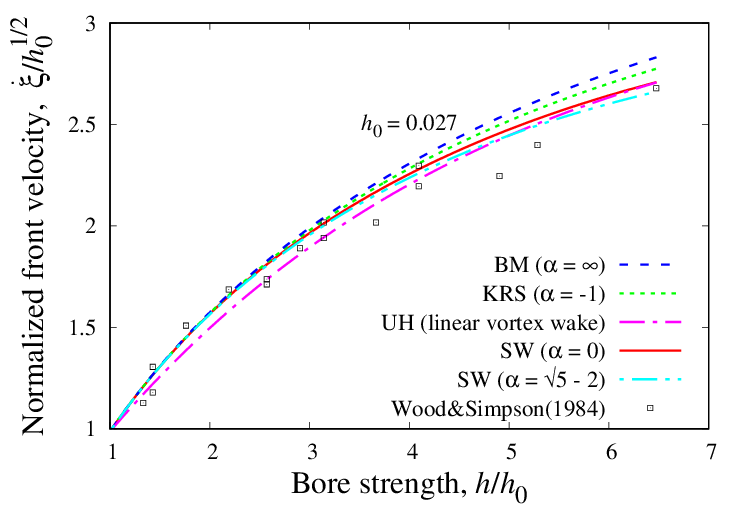}\put(-195,145){(\textit{a})}\includegraphics[width=0.5\columnwidth]{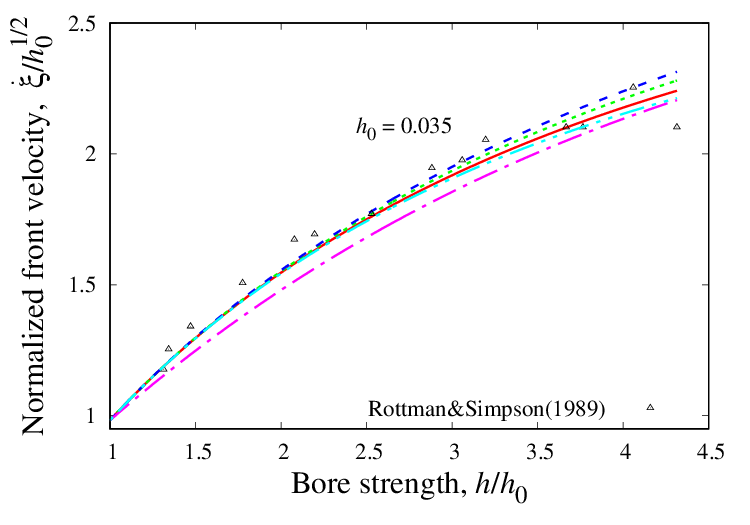}\put(-195,145){(\textit{b})}\\
\includegraphics[width=0.5\columnwidth]{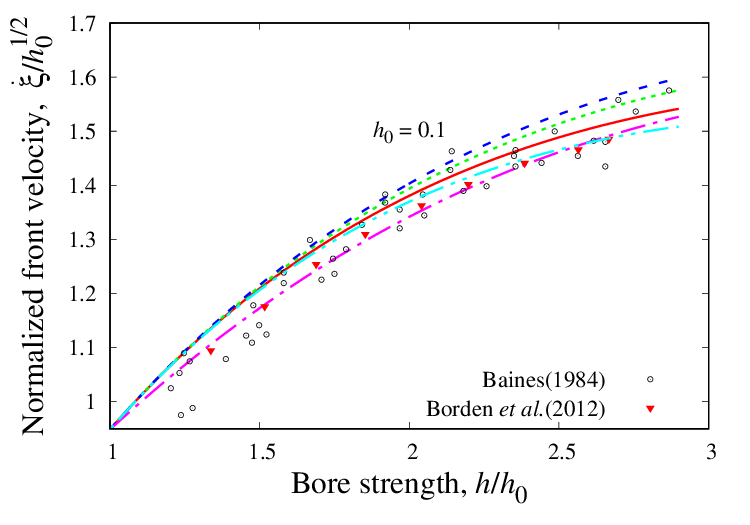}\put(-195,145){(\textit{c})}\includegraphics[width=0.5\columnwidth]{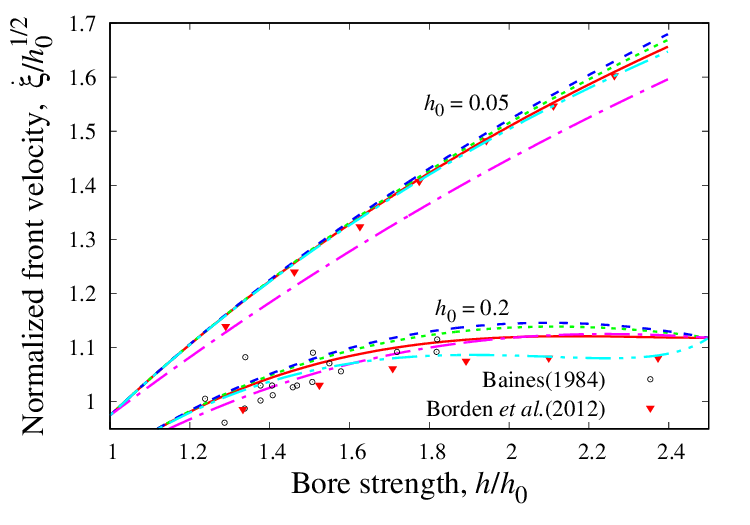}\put(-195,145){(\textit{d})}\caption{\label{fig:bore}The front velocity $\dot{\xi}/\sqrt{h_{0}}$ normalized
with the dimensionless depth of the bottom layer $h_{0}$ ahead the
bore versus the bore strength $h/h_{0}$ for $h_{0}=0.027,0.035,0.05,0.1,0.2:$
comparison of the SW theory $(\text{SW, }\alpha=0,\sqrt{5}-2$) with
KRS \citep{Klemp1997}, BM \citep{Borden2013}, UH \citep{Ungarish2018}
models as well as with the experimental results of \citet{Wood1984},
\citet{Rottman1983}, \citet{Baines1984} and the numerical results
of \citet{Borden2012}.}
\end{figure}

The aforementioned models are compared in Fig. \ref{fig:bore} with
the experimental results of \citet{Wood1984}, \citet{Rottman1983}
and \citet{Baines1984} as well as with the two-dimensional numerical
results of \citet{Borden2012}. Note that the ratio of densities $s=\rho^{-}/\rho^{+}=0.79$
used by \citet{Baines1984} is somewhat lower than $s=1$ assumed
in the Boussinesq approximation. Nevertheless, there is no noticeable
deviation of the experimental results from the Boussinesq approximation
when the average density is used as the characteristic value. For
consistency with previous studies, all front velocities are rescaled
with $\sqrt{h_{0}},$ which is the dimensionless velocity of small-amplitude
long interfacial waves when the depth of the bottom layer ahead of
the bore is small $(h_{0}\ll1).$ The front velocities normalized
in this way are plotted in Fig. \ref{fig:bore} against the bore strength
$h/h_{0},$ where $h$ is the upstream interface height. With this
normalization, we have $\dot{\xi}/\sqrt{h_{0}}\rightarrow1$ when
the downstream layer is thin $(h_{0}\rightarrow0)$ and the bore is
weak $(h/h_{0}\rightarrow1).$ All models can be seen to converge
to this essentially linear limit. Although the predicted front velocities
start to diverge at larger bore strengths, the divergence remains
small relative to the scatter in the experimental data. All front
velocities converge again, as for the gravity current velocity in
Fig. \ref{fig:gc}, when the interface height approaches the mid-plane
$h=0.5.$ The same front velocity produced by all models implies that
all underlying conservation laws are satisfied simultaneously in this
particular case. As noted before, this is the case for all bores with
the upstream interface located at the channel mid-height.

The SW front velocity (\ref{eq:xid}) for $\alpha=0$ is seen to approach
this limit tangentially. This is due to the distinctive feature of
this model, which yields $\frac{\mathrm{d}\dot{\xi}}{\mathrm{d}h}=0$
at $h=0.5,$ whereas all other models have $\frac{\mathrm{d}\dot{\xi}}{\mathrm{d}h}<0$
at this point. Numerical results for $h_{0}=0.2$ can be seen in Fig.
\ref{fig:bore}(d) to reproduce this nearly monotonous variation predicted
by the SW model with $\alpha=0$, though at sightly lower propagation
velocities. This difference, which is usually attributed to the turbulent
mixing between the layers, may also be due to viscous loss of momentum
at the rigid top and bottom boundaries. Viscous effects are assumed
to be negligible in the SW model but could be significant at the relatively
small Reynolds number $\mathit{Re}=3500$ used in the numerical simulation
by \citet{Borden2012}.

A remarkably better agreement with numerical results is produced by
the SW model with $\alpha=\alpha_{c}=\sqrt{5}-2.$ As discussed above
Eq. (\ref{eq:acrt}), physical considerations suggest that gravity
currents exceeding a certain critical height, which depends on $\alpha$
and is defined by the maximal propagation velocity for that $\alpha$,
are unstable. The largest height a stable gravity current can have
is attained at $\alpha=\alpha_{c}.$ This maximization of the front
height may be a dynamical mechanism behind the selection of $\alpha_{c}.$

\section{\label{sec:bores}Bores trailing gravity currents}

In the previous section, we showed that bores which propagate into
the quiescent downstream state can be described using the SW jump
conditions (\ref{eq:jmp1},\ref{eq:jmp2}). These conditions can be
applied also to more complex jump configurations as it is demonstrated
in this section for bores which can form on the top of gravity currents.
The presence of such bores can make the front of gravity current shallower
than the far upstream state. This may explain why experimentally observed
propagation velocities of shallow gravity currents are lower than
the theoretical predictions based on the upstream height.

Now, instead of the quiescent fluid layer shown in Fig. \ref{fig:sktch2},
the downstream state is assumed to be a gravity current with the interface
height $\eta_{0}$ and the shear velocity $\vartheta_{+}=\vartheta_{0}(\eta_{0})$
which is defined by Eq. (\ref{eq:vgc}) with $\eta_{0}$ substituted
for $\eta$. As before, the upstream height $\eta_{-}=\eta_{1}$
and the associated shear velocity $\vartheta_{-}=\vartheta_{1}$ need
to be found by solving Eqs. (\ref{eq:jmp1}, \ref{eq:jmp2}). For
$\alpha=0,$ using the computer algebra software Mathematica \citep{Wolfram2003},
we obtain: 
\begin{equation}
\vartheta_{1}^{\pm}(\eta_{1},\eta_{0})=\frac{(\eta_{1}+\eta_{0})(1-\eta_{1}\eta_{0})\vartheta_{0}\pm(\eta_{1}-\eta_{0})\gamma}{\eta_{1}+\eta_{0}-3\eta_{1}^{2}\eta_{0}+\eta_{1}^{3}},\label{eq:v1}
\end{equation}
where $\gamma^{2}=\eta^{2}\left(\left(4\eta_{1}^{2}-1\right)\vartheta_{0}^{2}-3\eta_{1}^{2}+1\right)-2\eta_{0}\eta_{1}\left(\eta_{0}^{2}+\vartheta_{0}^{2}-1\right)+\eta_{1}^{2}\left(\eta_{1}^{2}-\vartheta_{0}^{2}+1\right)$
and the plus and minus signs denote two possible branches of the solution.
A similar but somewhat longer solution can be obtained also for general
$\alpha.$ For $\eta_{1}=\eta,$ Eq. (\ref{eq:v1}) reduces to $\vartheta_{1}^{\pm}=\vartheta_{0},$
which corresponds to a uniform gravity current of the height $\eta_{0}.$

Let us first consider a shallow gravity current of the depth $h_{0}=(1+\eta_{0})/2\rightarrow0$
and assume the upstream state to be of a comparably small depth: 
\begin{equation}
h_{1}=(1+\eta_{1})/2=\kappa h_{0},\label{eq:h1}
\end{equation}
where $\kappa=h_{1}/h_{0}=O(1).$ In this limit, the propagation velocity
(\ref{eq:xid}) for the shear velocity $\vartheta$ defined by Eq.
(\ref{eq:v1}) becomes independent of $\alpha:$ 
\begin{equation}
\dot{\xi}_{1}=\left(1\mp\frac{\kappa}{\sqrt{\kappa+1}}\right)\dot{\xi}_{0},\label{eq:xid1}
\end{equation}
where $\dot{\xi}_{0}=\sqrt{2h_{0}}$ is the velocity of gravity current
for $h_{0}\rightarrow0.$ As seen, only the velocity defined by the
minus sign is physically feasible, i.e., $\dot{\xi}_{1}\le\dot{\xi}_{0}.$
Note that the bore velocity drops with the increase of its relative
height $\kappa$ and turns zero at 
\[
\kappa_{0}=\left(1+\sqrt{5}\right)/2.
\]
At this height ratio, the bore turns into a stationary hydraulic jump.
On the other hand, the energy balance defined by Eq. (\ref{eq:deps}),
which reduces to 
\begin{equation}
\dot{\varepsilon}_{1}=\frac{4\sqrt{2}(\kappa-1)^{3}\left(\kappa-\sqrt{\kappa+1}\right)}{(\kappa+1)^{3/2}}h_{0}^{5/2},\label{eq:eps1}
\end{equation}
indicates that the bore satisfies the energy constraint $\dot{\varepsilon}_{1}\le0$
only for $1\le\kappa\le\kappa_{0}.$ It means that a shallow gravity
current can be connected to an upstream state which is up to a factor
of $\kappa_{0}$ taller. Taking the maximum possible upstream depth
as the effective gravity current height reduces the normalized front
speed from $\sqrt{2}$ to $\dot{\xi}/\sqrt{h_{1}}=2/\sqrt{1+\sqrt{5}}\approx1.11.$
It is interesting to note that for a fixed upstream height $(\kappa h_{0}=\text{const}),$
$\dot{\varepsilon}_{1}$ defined by Eq. (\ref{eq:eps1}) attains minimum
at $\kappa_{c}\approx1.414.$ This yields the normalized front speed
$\dot{\xi}/\sqrt{h_{1}}\approx1.189,$ which is very close to the
empirical value of $1.19$ found for $h_{1}<0.075$ by \citet{Huppert1980}.

\begin{figure}
\centering{}\includegraphics[bb=90bp 50bp 360bp 302bp,clip,width=0.5\columnwidth]{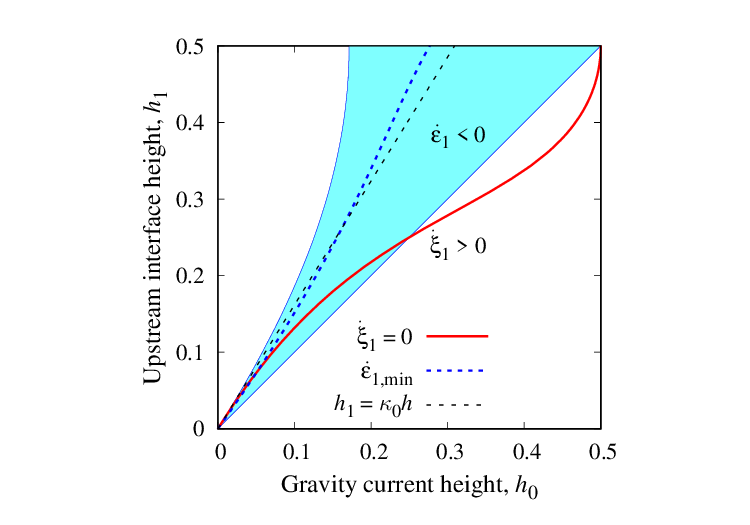}\put(-210,200){(\textit{a})}\includegraphics[bb=90bp 50bp 360bp 302bp,clip,width=0.5\columnwidth]{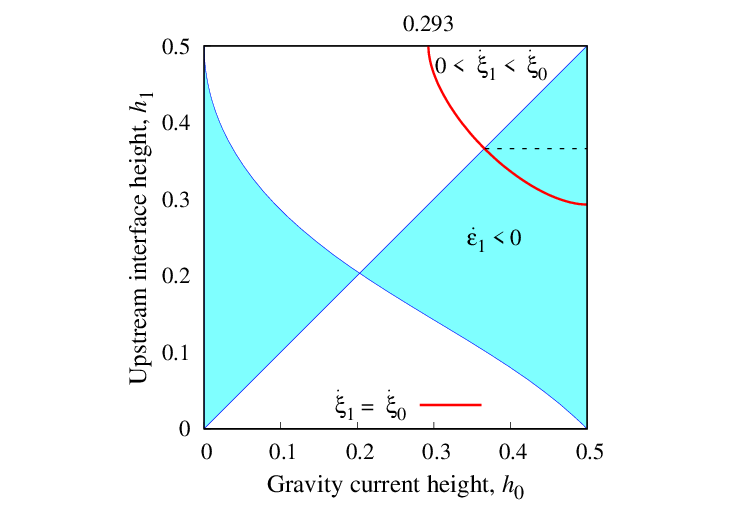}\put(-210,200){(\textit{b})}
\caption{\label{fig:gcf}The upstream interface height $h_{1}$ permitted by
the energy balance constraint $\dot{\varepsilon}_{1}\le0$ defined
by Eq. (\ref{eq:deps}) versus the gravity current height $h_{0}$
for the positive (\emph{a}) and the negative (\emph{b}) branches of
Eq. (\ref{eq:v1}) $(\alpha=0).$ The dotted lines show the upper
limit for shallow gravity currents $(h_{1}=\kappa_{0}h_{0})$ and
the depth at which the energy loss attains maximum ($\dot{\varepsilon}_{1,\min})$;
the solid lines show the interface height at which the upstream jump
becomes stationary $(\dot{\xi}_{1}=0)$ (\emph{a}) and moving at the
same speed as the gravity current $\dot{(\xi}_{1}=\dot{\xi}_{0})$
(\emph{b}).}
\end{figure}

The exact solution of Eq. (\ref{eq:deps}) plotted in Fig. \ref{fig:gcf}
shows that the possible upstream depth for deeper gravity currents
is larger than that predicted by the linear relationship (\ref{eq:h1})
for shallow currents. Nevertheless, the height of maximal energy loss
for a fixed upstream depth remains relatively close to this line also
for deeper currents. 

First, let us consider the possibility of a shallower gravity current
forming at the front of a supercritical current with $h\ge h_{c}.$
This corresponds to a relatively deep leading gravity current connected
to a taller upstream state via single bore which moves in the same
direction as the gravity current. As seen in Fig. \ref{fig:gcf}(a),
if the leading gravity current is sufficiently shallow, the bores
described by the positive branch of Eq. (\ref{eq:v1}) can move downstream
only. The bores that can trail deep gravity currents are described
by the negative branch of Eq. (\ref{eq:v1}). The corresponding height
of gravity current and that of bore, which are admitted by the energy
balance constraint (\ref{eq:deps}), are shown by the filled-in region
on Fig. \ref{fig:gcf}(b). The solid line delimits the range of bores
whose speed of propagation does not exceed that of the leading gravity
current. The interface heights that satisfy both constraints are located
in the upper right corner below the diagonal line $(h_{1}\le h_{0})$
on Fig. \ref{fig:gcf}(b). This corresponds to an upstream state which
is shallower than the leading gravity current head. The latter, in
turn, has to be taller than the minimal depth defined by the intersection
of the diagonal with the solid line along which $\dot{\xi}_{1}=\dot{\xi}_{0}.$
This point coincides with the critical depth $h_{c}$ at which the
leading front speed attains maximum and becomes unstable at $h>h_{c}.$

\begin{figure}
\centering{}\includegraphics[width=0.5\columnwidth]{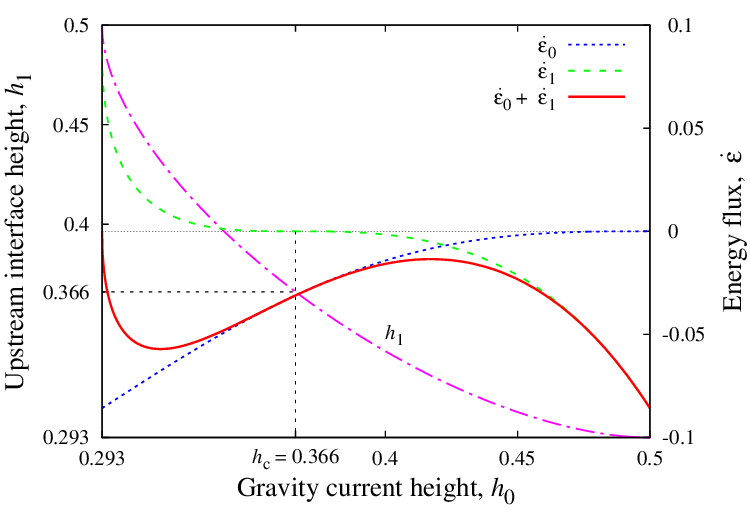} \caption{\label{fig:jmp}The bore height $h_{1}$ versus the depth of gravity
current $h_{0}$ at which both move at the same speed $\dot{\xi}_{1}=\dot{\xi}_{0}$
along with the energy balance defined by Eq. (\ref{eq:deps}) for
the gravity current $(\dot{\varepsilon}_{0})$ and the bore $(\dot{\varepsilon}_{1})$.}
\end{figure}

It may also be seen in Fig. \ref{fig:gcf}(b) that stable gravity
currents, which are located in the region above the diagonal with
$h_{0}<h_{c}<h_{1},$ give rise only to energy-generating bores. The
latter are unphysical when considered as separate entities. Hypothetically,
such bores can exist together with a leading gravity current provided
that no total energy is generated by the system of the two coupled
jumps. However, for such a coupled system to form, the bore has to
move at the same speed as the leading front, i.e., $\dot{\xi}_{1}=\dot{\xi}_{0}$.
The jump heights which satisfy this constraint are shown by the solid
line in \ref{fig:gcf}(b). As seen in Fig. \ref{fig:jmp}, although
the bore generates energy if $h_{0}<h_{c},$ more energy is dissipated
by the leading gravity current front as long as the upstream state
is shallower than the channel mid-height $h_{1}=0.5$. This critical
upstream depth produces a leading gravity current of the lowest possible
depth $h_{0}=1-1/\sqrt{2}\approx0.293$ for $\alpha=0.$ The corresponding
depths for $\alpha=-1$ and $\alpha=\infty$ are $(5-\sqrt{17})/4\approx0.219$
and $(3-\sqrt{5})/4\approx0.191,$ respectively. The decrease of the
upstream depth causes the leading gravity current depth rise until
both become equal at $h_{c}$ which is defined by Eq. (\ref{eq:hc}).
At this point, the bore height becomes small relative to the underlying
gravity current. Thus, its velocity of propagation approaches that
of small-amplitude disturbances. This velocity equals the characteristic
wave speed $\lambda_{+}$ defined by Eq. (\ref{eq:Cpm}). It explains
why $\lambda_{+}$ coincides with the maximum of $\dot{\xi}$ exactly
rather than just approximately as speculated by \citet{Baines2016}.
It also reveals the duality of the propagation velocity for gravity
currents of a supercritical depth $h_{1}\ge h_{c}.$ Firstly, $\dot{\xi}=\dot{\xi}(h_{1})$
defines the velocity of a gravity current of depth $h_{1}.$ Secondly,
this velocity is equal to the velocity of a bore with the same upstream
height which trails the gravity current of a subcritical depth $h_{0}<h_{c}$
propagating at the same speed $\dot{\xi}=\dot{\xi}(h_{0})$. 

\section{\label{sec:Sum}Summary and Conclusion}

In this paper, we derived a locally conservative SW momentum equation
for the two-layer system bounded by a rigid lid. The equation contains
a free dimensionless parameter $\alpha,$ which defines the relative
contribution of each layer to the pressure gradient along the interface.
This equation can describe strong internal bores and gravity currents
in a self-contained way without invoking external front conditions.
So far such closure conditions were presumed to be indispensable and
derived using various control-volume methods. The momentum equation
(\ref{eq:U-gen}) was obtained for two fluids with arbitrary density
difference by using a linear combination of the basic SW equations
(\ref{eq:upm}), which describe the conservation of irrotationality
in each layer, to eliminate the pressure gradient along the interface.
Applying the Boussinesq approximation to fluids with a small density
difference, the momentum equation reduced to Eq. (\ref{eq:uh}). Finally,
using the velocity and height differentials as dynamical variables,
this equation was written in the dimensionless form as Eq. (\ref{eq:gen}).

The appearance of the free parameter $\alpha$ in the general form
of the momentum equation is closely related with the presence of the
external length scale, the total height $H,$ in the considered two-layer
system. Namely, using the fact that the circulation conservation equation
(\ref{eq:u}) is a one-order-lower SW conservation law in terms of
the height than the momentum equation (\ref{eq:U}), we can multiply
the former with a coefficient $\propto H$ and then add it to the
latter. Such an operation is formally permitted because both equations
have the same physical units. This leads to the generalized momentum
conservation equation containing $\alpha$ and the associated component
of the circulation conservation law. The $\alpha$-term in the generalized
momentum equation (\ref{eq:U-gen}) vanishes if this equation is satisfied
simultaneously with the circulation conservation equation (\ref{eq:u}).
This, however, is the case for smooth waves but not in general for
hydraulic jumps.

Dimensional as well as physical considerations suggest that if $\alpha$
is constant, as it is required by the equivalence of momentum and
circulation conservation laws for smooth waves, it can depend only
on the ratio of densities, which is the sole dimensionless parameter
in this problem. As the dynamical pressure produced by the flow is
proportional to the density of fluid, for nearly equal densities,
both layers can be expected to affect the interfacial pressure gradient
with approximately equal weight coefficients. This corresponds to
$\alpha\approx0.$

The jump propagation velocity (\ref{eq:xid}) which results from the
Rankine-Hugoniot jump conditions (\ref{eq:jmp1}, \ref{eq:jmp2})
for the mass conservation equation (\ref{eq:vlm}) and the momentum
equation (\ref{eq:gen}) with $\alpha=0$ was compared with the predictions
of a number of previous models as well as with the numerical and experimental
results. A good agreement was found with the available data including
those for moderately non-Boussinesq fluids. The propagation velocities
resulting from the SW model with $\alpha=0$ appear generally closer
to the numerical results than the those predicted by the previous
models. We note that a mathematically equivalent result is produced
by the so-called vortex-sheet model of \citet{Ungarish2018} using
the head loss ratio $\lambda=-1.$ However, negative head loss ratios
are not compatible with the basic control volume assumptions. This
is discussed in more detail below as well as in the Appendix.

A particularly good agreement with high-accuracy numerical results
for both gravity currents and bores is produced by $\alpha_{c}=\sqrt{5}-2.$
At this exceptional value, the gravity current speed becomes a monotonically
increasing function of its depth. At all other $\alpha,$ the gravity
current speed attains a maximum at certain critical depth which depends
on $\alpha.$ Simple physical considerations suggest that gravity
currents of supercritical depth are unstable. The largest front height
that a stable gravity current can have is attained at $\alpha_{c}$.
We hypothesize that this value may be selected dynamically when an
unstable gravity current collapses from a supercritical height to
the largest possible stable height. Alternatively, the instability
may cause the front height to increase producing an elevated gravity
current head which may rise up to the channel mid-height as it is
shown by the exact analytical solution as well as observed in 2D numerical
simulation of the partial lock exchange problem \citep{Politis2020,Khodkar2017}.

We also showed that the classical front condition for gravity currents
obtained by \citet{Benjamin1968} as well as its generalization to
internal bores by \citet[(KRS)][]{Klemp1997} are reproduced by the
SW momentum equation with $\alpha=-1.$ This value describes the interfacial
pressure gradient which is determined entirely by the bottom layer.
Similarly, the alternative front condition proposed earlier by \citet{Wood1984},
which is known to yield a generally poorer agreement with experimental
results than the KRS model and to break down for shallow gravity currents,
is reproduced by the SW model with $\alpha=1.$ This value describes
the interfacial pressure gradient determined solely by the top layer.

According to the traditional interpretation, in the WS model, the
energy is conserved in the top layer and lost only in the bottom layer,
while in the KRS model, it is the other way round \citet{Li1998}.
argue that, in general, energy can be lost simultaneously in both
layers. If so, then the WS and KRS models would represent two limiting
cases and yield, respectively, the upper and lower bound on the bore
velocity. It is important to note that the bore velocities resulting
from the SW model with $\alpha=0$ are lower than those predicted
not only by the WS but also by the KRS model. This is because only
the total energy for both layers together is defined in the SW framework.
It is important stress that this is a rigorous mathematical fact rather
than a deficiency of the SW model, in which one layer can gain energy
from the other as long as the total energy does not increase across
the jump. The latter condition, which is defined by Eq. (\ref{eq:deps}),
determines in which direction bore can propagate. In the conventional
control-volume approach, the energy exchange between the layers is
absent because the interface is assumed to be stationary in the co-moving
frame of reference. Therefore, in a self-consistent control-volume
approach, neither layer can gain energy and hence neither head loss
can be negative \citep{Li1998}.

There is no such a constraint in the SW framework, where the pressure
drop across the hydraulic jump in each layer follows from the conservation
of irrotationality (\ref{eq:upm}). If the interface in the hydraulic
jump is stationary in the co-moving frame of reference, as it is commonly
assumed in the control-volume approach, the same pressure drop would
result also from the energy conservation in each layer which is described
by Eq. (\ref{eq:Ep}). Such an assumption, however, is unphysical
and leads to a paradox which is illustrated below by the circulation
conservation condition. It is important to note that, in the hydrostatic
SW approximation, the solution that is stationary (in the co-moving
frame of reference) cannot be continuous (and the other way round)
\citep{Whitham1974}. The control-volume analysis being an integral
approach misses this essential mathematical subtlety.

The circulation conservation condition is effectively based on the
assumption that the pressure drop across the jump is the same in both
layers. In the control-volume approach, the corresponding pressure
drops are assumed to follow from the conservation of energy in each
layer. The paradox is that the assumed conservation of energy across
the jump in each layer separately does not guarantee the conservation
of energy in both layers together. It is because, in the hydrostatic
SW approximation as well as in the control-volume approach, the energy
across the jump cannot in general be conserved simultaneously with
the mass and circulation. This paradox appears also in the inviscid
limit of the vortex-sheet model of \citep{Ungarish2018} who attribute
it to the inadequacy of the inviscid approximation. From the SW perspective
described above, this paradox is due to the inadequacy of the energy
conservation assumption in separate fluid layers.

Similar to the energy also the circulation cannot in general be conserved
simultaneously with the momentum and mass. In the SW framework, this
is known as the Rankine-Hugoniot deficit \citep{Kalisch2017}. The
conservation of a given quantity depends on the physical mechanisms
which are outside the scope of the SW approximation but could become
relevant in the hydraulic jumps. Such mechanisms are, for example,
the viscous dissipation and turbulence, which can account for the
loss of energy and the generation of circulation in strong bores.
But there is no analogous physical mechanism which could disrupt the
momentum balance in highly inertial flows. Therefore, the momentum
conservation, notwithstanding its inherent ambiguity, appears physically
more relevant than the conservation of circulation and energy.

The main advancement of the proposed theory is the realization that
internal bores and gravity currents can be described using hydrostatic
SW approximation which was previously thought impossible. The proposed
SW theory provides not only a mathematically consistent and rational
alternative to the conventional control-volume approach but also a
self-contained framework for numerical modeling of strong internal
bores and gravity currents in the two-layer systems bounded by a rigid
lid. A canonical example is the lock-exchange problem which can be
solved analytically by the method of characteristics and used as a
benchmark to validate numerical solution of the locally conservative
SW water equations derived in this study \citep{Politis2020}.

\section*{Appendix}

This section presents an extended discussion of the key differences
between the proposed SW theory and two alternative approaches \citep{Borden2013,Ungarish2018}.
The first is that of \citet{Borden2013} who argue that since the
vorticity front condition can be decoupled from the pressure drop,
the latter can be determined, if required, independently from the
former by substituting the jump parameters that satisfy this front
condition into the momentum balance condition. So both conditions
would be satisfied simultaneously along with the conservation of mass.
This, however, is at odds not only with the SW theory but also with
the inviscid limit of the vortex-sheet model of \citet{Ungarish2018}
which is physically equivalent to the model of \citet{Borden2013}.
The fact that the vorticity equation, which \citet{Borden2013} integrate
over the control volume enclosing the jump, contains no pressure,
does not decouple the latter from the vorticity front condition. It
is important to note that the 2D integral of the vorticity transport
equation considered by \citet{Borden2013} is mathematically equivalent
by Stokes' theorem to the circulation of Euler equation along the
boundary of the control volume \citep{Ungarish2018}. Therefore, it
is not mathematically consistent to accept one approach and dismiss
the other as done, for example, by \citet{Khodkar2017} with respect
to Bernoulli's law which underlies the vorticity front condition.
The pressure drop across a circulation-conserving hydraulic jump is
indeed defined by the second approach. This pressure drop, however,
is not necessarily the same as that satisfying the momentum balance
condition. Thus, the latter cannot be satisfied together with the
circulation conservation unless the model is substantially extended
by introducing two additional phenomenological parameters, the so-called
head losses, which are supposed to account for the effect of viscous
dissipation on the pressure drop \citep{Ungarish2018}.

\citet{Borden2013} also argue that their vorticity front condition
must satisfy the momentum balance because so does the vorticity equation
it is derived from. The SW equations, which follow from the mathematically
equivalent Euler equation, demonstrate that the equivalence of the
associated conservation laws relies on the continuity of the velocity
and pressure fields. Continuity is required only by the PDEs, which
describe local conservation laws, but not by the integral conservation
laws, which are actually employed for the derivation of approximate
front conditions using the control-volume method.

The second alternative approach is the vortex-sheet model of \citet{Ungarish2018}.
As noted before, the front speed resulting from the Rankine-Hugoniot
jump conditions of the generalized SW momentum and mass conservation
equations is mathematically equivalent to that resulting from the
vortex-sheet model when the ratio of head losses is expressed as $\lambda=\frac{\alpha-1}{\alpha+1}.$
This equivalence is intriguing because the head loss is a phenomenological
quantity which is introduced in the control-volume approach to account
for the viscous energy losses in each layer. Such a quantity, however,
is completely outside the scope of the SW theory, which neither requires
nor allows energy dissipation to be attributed to a particular layer.
In contrast to the head loss, the free parameter $\alpha$ which features
in the SW theory is not a phenomenological parameter modeling something
outside the SW approximation. As a matter of fact, $\alpha$ is an
intrinsic parameter of the two-layer SW system bounded by rigid lid
where it emerges due to the inherent non-uniqueness of the SW momentum
equation.

It has be noted that \citet{Ungarish2018} substantially extend the
original meaning of head loss by associating it with the viscous flux
of vorticity through the adjacent horizontal boundary. The inferred
direction of this flux is then used to conclude that head loss must
be positive in the top layer and negative in the bottom layer. This
may be in agreement with numerical simulations but, as noted in the
conclusion, it is not consistent with the stationarity of the interface,
which is assumed in the control-volume approach and requires the head
losses in both layers to be non-negative \citep{Li1998}.

There are two more questionable assumptions underlying the vortex-sheet
model. First, \citet{Ungarish2018} consider a viscous fluid but substitute
the physical no-slip conditions at the solid top and bottom walls
with artificial free-slip conditions. This substitution is essential
for the control-volume method to be applicable. On the other hand,
the exact boundary conditions at the top and bottom walls may not
be that relevant as indicated by the numerical simulations \citealp{Haertel2000}.

However, the main premise of the vortex-sheet model is that the balance
of vorticity, which is generated by the the baroclinic torque at the
interface and then advected by the flow along the layer, can significantly
be affected by the viscous vorticity fluxes through the top and bottom
boundaries at which a zero vorticity is imposed by the free-slip boundary
conditions. There are two problems with this assumption in relation
to highly inertial flows which are typical for turbulent internal
bores and gravity currents. Firstly, at high Reynolds numbers, the
vorticity generated by the baroclinic torque at the interface is predominantly
advected downstream, especially in the upper layer. Thus, the vorticity
remains confined within a relatively thin layer forming along the
interface, as it is assumed in the more refined vortex-wake model
\citep{Ungarish2018}. As a result, the vorticity is advected out
of the control volume through the downstream side boundary well before
it reaches the top wall. Besides that the free-slip boundary condition
allows the flow along the top wall to remain irrotational far downstream
from the control volume. Consequently, there can be no significant
viscous flux of vorticity through the upper boundary of the control
volume. This is confirmed by the numerical results of \citet{Nasr-Azadani2015},
who find no significant difference between the free-slip (zero vorticity)
and zero-vorticity-flux boundary conditions.

The same argument as above holds also for the bottom layer except
the gravity currents for which this layer is nearly quiescent in the
co-moving frame of reference. Although the viscous vorticity flux
can be relatively significant in this case, one needs to take into
account the characteristic vorticity diffusion time $\sim H^{2}/\nu$
which is necessary for a steady vorticity flux to develop. This time
is $\sim\mathit{Re}\gtrsim10^{4}$ relative to the time $\sim H/U$
taken by fluid to flow over a horizontal distance comparable to the
layer depth. It means that viscous vorticity fluxes cannot develop
on the time scale relevant to the propagation of turbulent internal
bores and gravity currents. In addition, it should be noted that there
is no viscous dissipation along at the boundaries when free-slip is
assumed. This makes the association of head losses with viscous vorticity
fluxes problematic.

In addition, the fact that only the ratio of head losses appears in
the general vortex-sheet front condition implies that these two phenomenological
quantities are not mutually independent. Firstly, they are related
to each other through the pressure balance condition at the interface.
This condition is equivalent to the pressure compatibility condition
used by \citet{Ungarish2018}. Secondly, in the hydrostatic approximation,
the head losses along the top and bottom boundaries follow directly
from the pressure drop along the interface. The latter is the key
quantity in the SW model where it is defined using single parameter
$\alpha$ while the head losses may be regarded as secondary quantities
depending on $\alpha$.

\section*{Acknowledgments}

Numerous stimulating discussions with G. Politis are gratefully acknowledged.
I would also like to thank P. Milewski and M. Ungarish for bringing
their recent papers to my attention.

\bibliographystyle{plainnat}
\bibliography{/home/priede/work/swe2d1l/ref/shwt}

\end{document}